 \documentclass[smallabstract,smallcaptions]{dccpaper}

\usepackage{epsfig}
\usepackage{amsmath}
\usepackage{amssymb}
\usepackage{color}
\usepackage{url}

\newlength{\figurewidth}
\newlength{\smallfigurewidth}

\usepackage{booktabs}
\begin{document}

\title
{\large
\textbf{InterGS+: Inter-Frame Gaussian Splatting Compression with Geometry-Aware Transformer Prediction and Adaptive Residual Coding}
}

\author{%
Muhammad Talha$^{\ast}$, Zhu Li$^{\ast}$, and Joel Jung$^{\dag}$\\[0.5em]
{\small\begin{minipage}{\linewidth}\begin{center}
\begin{tabular}{ccc}
$^{\ast}$University of Missouri-Kansas City & \hspace*{0.5in} & $^{\dag}$Qualcomm, Inc. \\
% Street Address One && Street Address Two \\
% City, State, ZIP, Country && City, State, ZIP, Country\\
\url{mtcf@umkc.edu} && \url{joeljung@qti.qualcomm.com}
\\
\url{zhu.li@ieee.org} &&
\end{tabular}
\end{center}\end{minipage}}
}

\maketitle
\thispagestyle{empty}
\begingroup
\renewcommand{\thefootnote}{}
\footnotetext{This work is supported by an award from NSF RINGS Project [CNS-2148382], and a gift grant from Qualcomm. Corresponding Author: Muhammad Talha (mtgcf@umkc.edu)}
\endgroup

\begin{abstract}
Dynamic 3D Gaussian Splatting (3DGS) provides real-time photorealistic
rendering, but representing each frame as irregular Gaussian primitives with
high-dimensional attributes creates substantial storage and transmission cost.
Exploiting temporal redundancy is challenging because Gaussian primitives do
not preserve fixed identities across frames. We propose \emph{InterGS+}, a
lightweight post-hoc inter-frame codec that jointly improves temporal prediction
and residual representation. Current-frame geometry is coded with G-PCC and
reconstructed before prediction, enabling consistent temporal neighborhoods at
encoder and decoder. Attributes are predicted using lightweight KNN and
bilateral predictors together with a geometry-aware transformer for
spherical-harmonic (SH) appearance, with per-Gaussian mode selection. Residual
coding is adapted to each attribute: SH innovations are coded in a shared PCA
domain using vector quantization with refinement, opacity and scale in activated
gamma-companded domains, and rotation as canonicalized quaternion innovations
using vector quantization and refinement. On MPEG dynamic sequences, InterGS+
achieves average BD-rate reductions of $45.48\%$ over G-PCC Amd.~1 and
$7.78\%$ over InterGS-Lite, with corresponding BD-PSNR gains of $9.98$ dB and
$1.10$ dB. Average decoding complexity is $0.049$ ms/Gaussian, demonstrating
efficient streamable dynamic 3DGS compression.
\end{abstract}

\vspace{-7mm}
\section{Introduction \& Related Work}
\label{sec:intro}
\vspace{-3mm}

3D Gaussian Splatting (3DGS) enables high-quality novel-view synthesis with
real-time rasterization~\cite{Kerbl2023}, but its explicit representation incurs
substantial storage and transmission cost. Each Gaussian carries geometry,
opacity, scale, rotation, and high-dimensional spherical-harmonic (SH)
appearance, all of which must evolve over time in dynamic scenes. Efficient
compression is therefore essential for practical free-viewpoint video and
volumetric-media applications.
Classical point-cloud codecs provide established tools for irregular geometry
and attributes, including MPEG G-PCC/V-PCC~\cite{Schwarz2019PCC} and spatial
transforms such as RAHT~\cite{Queiroz2016RAHT}; however, 3DGS introduces a
richer, more heterogeneous signal. Static Gaussian compression has explored pruning, quantization, compact
appearance models, predictive parameterization, spatial context modeling, and
learned entropy coding~\cite{Niedermayr2024,Lee2024Compact3DGS,
Fan2024LightGaussian,Liu2024CompGS,Chen2024HAC,Wang2024ContextGS,
Chen2025FCGS}. Other methods reorganize Gaussian attributes for conventional
video coding~\cite{Lee2025CodecGS}, while recent work investigates voxelized,
hierarchical, and video-codec-based representations~\cite{Fu2026VoxelGS,
Sridhara2026RALHE,Yang2026LGSCV}. MPEG has also begun studying interoperable
Gaussian Splat Coding~\cite{Bang2026MPEGGSC}. These approaches reduce
redundancy within individual Gaussian representations, but dynamic sequences
introduce additional redundancy over time. Exploiting it is difficult because
Gaussian primitives are irregular, unordered, and need not preserve identities
across frames. Existing dynamic methods use trajectory
simplification~\cite{Javed2025TC3DGS}, spatial-temporal primitive
prediction~\cite{Liu2025CompGSpp}, or learned group-of-frames motion and entropy
models~\cite{Zhang2026DFCGS}, but often remain coupled to a particular
trajectory, motion model, optimization process, or scene representation.

InterGS~\cite{Talha2025InterGS} introduced a \emph{post-hoc} framework that
predicts attributes of already generated Gaussian frames from geometrically
related primitives in the reconstructed reference. InterGS-Lite
~\cite{Talha2026InterGSLite} retained this representation-independent temporal
framework, using lightweight KNN/bilateral prediction and vector-quantized SH
residuals to exploit temporal redundancy directly in the Gaussian domain
without modifying the underlying scene representation or introducing a heavy
learned decoder. InterGS+ extends this same framework in two directions: it
introduces a learned geometry-aware SH predictor in a shared PCA domain, and
replaces raw non-SH residual formation with attribute-specific representations
for opacity, scale, and rotation.

% InterGS~\cite{Talha2025InterGS} instead introduced a \emph{post-hoc} framework that predicts attributes of already generated Gaussian frames from geometrically related primitives in the reconstructed reference. InterGS-Lite~\cite{Talha2026InterGSLite} retained this representation-independent temporal framework and improved SH residual coding through vector quantization, showing that temporal redundancy can be exploited directly in the Gaussian domain without modifying the underlying scene representation or introducing a heavy learned decoder.

Two limitations remain. First, the lightweight KNN and geometry-weighted predictors of InterGS-Lite do not fully exploit the structure of high-dimensional SH appearance, whose evolution can depend jointly on reference appearance and neighborhood geometry. Second, Gaussian attributes do not share a common parameter space: opacity and scale are interpreted after nonlinear renderer activations, while rotation has equivalent quaternion and scale--rotation parameterizations. Residuals formed without accounting for these properties can therefore preserve redundancy that is irrelevant to the rendered Gaussian.

InterGS+ addresses both limitations within the same causal post-hoc coding framework. SH appearance is projected into a fixed global PCA space shared by prediction and residual coding, and a lightweight geometry-aware transformer is added as an adaptive SH prediction candidate alongside the existing local predictors. For non-SH attributes, residual formation is matched to the representation itself: opacity and scale are coded after their native activations and power companding, while scale--rotation canonicalization and quaternion sign alignment remove equivalent parameterizations before rotation coding. Rotation innovations are then represented jointly through vector quantization followed by scalar refinement.

Taken together, these extensions strengthen both stages of predictive coding: richer geometry-aware SH prediction reduces temporal uncertainty, while representation-aware residual formation makes the remaining innovation more compressible without changing the underlying dynamic 3DGS representation.

\noindent\textbf{The main contributions are:}
\vspace{-3mm}
\begin{itemize}

    \item We introduce a fixed global PCA representation for SH appearance,
    reducing 48 coefficients to 38 components and providing a shared compact
    domain for transformer training, temporal prediction, and SH residual
    coding without per-sequence transform signaling.

    \vspace{-3mm}

    \item We introduce a lightweight \emph{geometry-aware SH transformer} for
    temporal prediction of high-dimensional SH appearance. Beyond standard
    content attention, the predictor explicitly encodes local geometry through
    neighbor-rank embeddings, learned geometry-offset embeddings, additive
    multi-frequency Fourier features, and pairwise relative-position bias in the
    attention logits. This allows the model to reason jointly over reference
    appearance and neighborhood structure, while per-Gaussian mode selection
    retains the simpler KNN and bilateral predictors where they remain more
    accurate.

    \vspace{-3mm}

    \item We introduce representation-aware residual formation for non-SH
    attributes: opacity and scale are coded after their native activations and
    power companding, while deterministic scale--rotation canonicalization,
    quaternion normalization, and sign alignment remove artificial
    parameterization differences before rotation residual formation.

    \vspace{-3mm}

    \item We introduce joint coding of the canonicalized four-dimensional
    rotation innovation using vector quantization followed by scalar
    refinement, preserving inter-component structure rather than coding
    quaternion residuals independently.

\end{itemize}

\vspace{-8mm}
\section{Proposed Method}
\label{sec:method}
\vspace{-4mm}

\subsection{Codec Overview and Temporal Reference Construction}
\label{subsec:overview}
\vspace{-2mm}

\begin{figure}[t]
    \centering
    \includegraphics[width=0.78\linewidth]
    {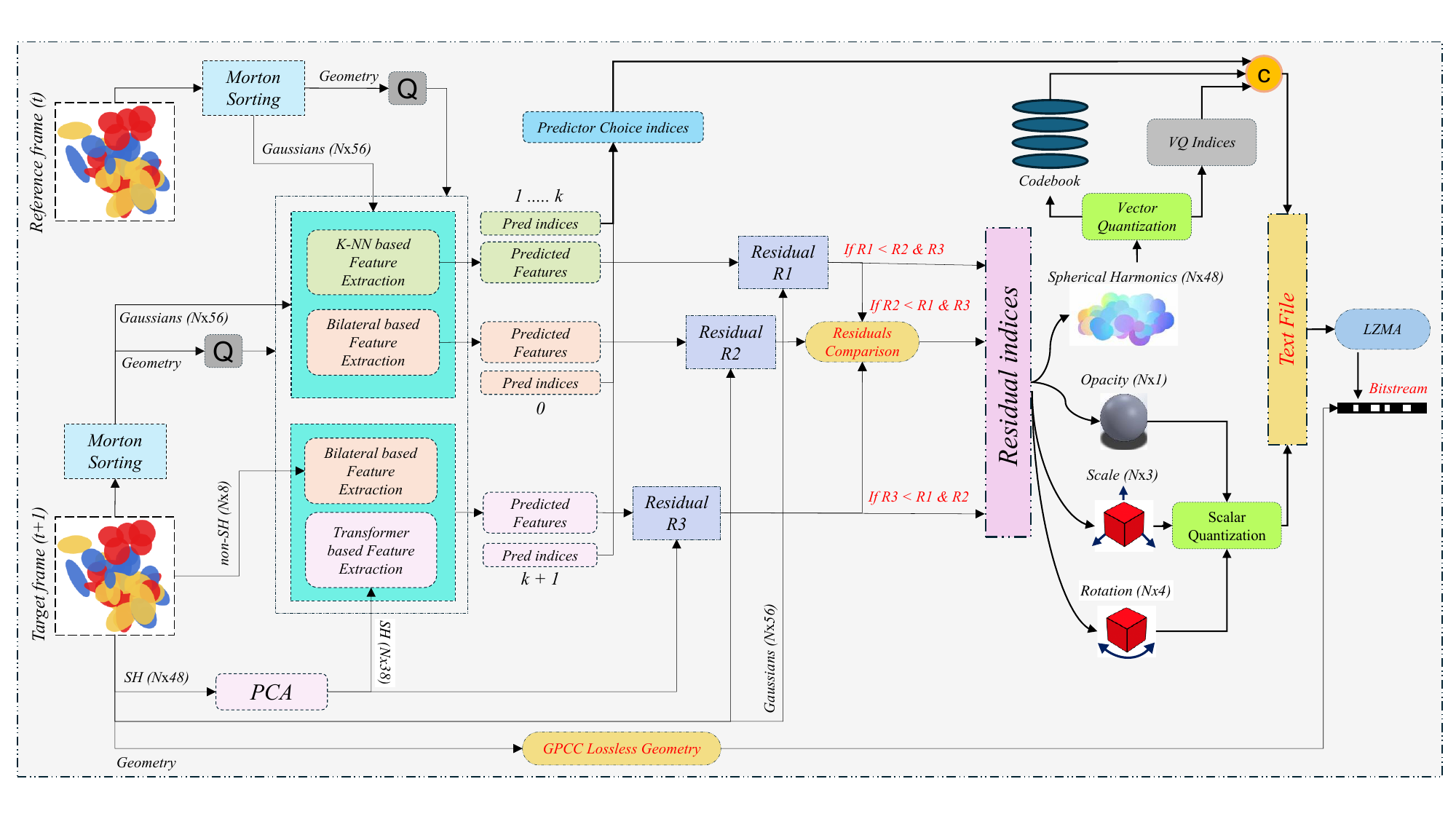}
    \vspace{-4mm}
    \caption{Overview of InterGS+. Quantized geometry establishes temporal
    support, a hybrid predictor removes temporal redundancy, and the remaining
    innovation is represented using attribute-adaptive residual coding.}
    \label{fig:main_architecture}
    \vspace{-6mm}
\end{figure}

% InterGS+ exploits temporal redundancy directly between reconstructed Gaussian
% frames while accounting for irregular geometry and heterogeneous attributes.
% Geometry establishes temporal support, prediction removes reference-predictable
% information, and the remaining innovation is coded in an attribute-specific domain.
% Figure~\ref{fig:main_architecture} summarizes the codec. Each Gaussian contains
% a 3-D center and a 56-D attribute vector consisting of 48 SH coefficients,
% opacity, three scale parameters, and a four-component rotation quaternion.
% The first frame of each GOP establishes the temporal reference, while each
% subsequent P-frame is predicted from the previously reconstructed Gaussian
% frame. This causal dependency requires no future-frame access and only a single
% reconstructed reference frame in memory; periodic I-frames provide random-access
% points at GOP boundaries. Although all frames use the same 18-bit geometry precision, their quantized
% coordinate systems may have different frame-dependent origins. Their integer
% coordinates therefore cannot be compared directly. Let $Q_t(\cdot)$ denote the
% geometry quantizer associated with frame $t$. The reconstructed reference
% geometry is mapped into the current-frame coordinate system before temporal
% neighbor search:
InterGS+ exploits temporal redundancy directly between reconstructed Gaussian
frames while accounting for irregular geometry and heterogeneous attributes:
SH is high-dimensional and correlated, opacity and scale use nonlinear
parameterizations, and rotation is constrained and non-unique. Geometry
establishes temporal support, prediction removes reference-predictable
information, and the remaining innovation is coded in an attribute-specific
domain. Figure~\ref{fig:main_architecture} summarizes the codec. Each Gaussian
contains a 3-D center and a 56-D attribute vector comprising 48 SH coefficients,
opacity, three scale parameters, and a four-component rotation quaternion.
The first frame of each GOP establishes the temporal reference; each subsequent
P-frame uses only the previously reconstructed Gaussian frame. This causal
structure requires no future-frame access and stores only one reconstructed
reference, while periodic I-frames provide random access at GOP boundaries. Although all frames use 18-bit geometry precision, frame-dependent quantizer
origins prevent direct comparison of their integer coordinates. Let
$Q_t(\cdot)$ denote the geometry quantizer for frame $t$; before temporal
neighbor search, reconstructed reference geometry is mapped into the
current-frame coordinate system:
\vspace{-4mm}
\begin{equation}
\mathbf q_{t,i}=Q_t(\mathbf x_{t,i}),
\qquad
\bar{\mathbf q}^{\,t}_{t-1,j}
=
Q_t\!\left(
Q^{-1}_{t-1}(\hat{\mathbf q}_{t-1,j})
\right).
\label{eq:geometry_align}
\vspace{-2mm}
\end{equation}

% Current geometry is coded with G-PCC as quantized integer geometry. Both
% current and reference Gaussians are stably Morton ordered for deterministic
% serialization; Morton order itself is not treated as temporal correspondence.
% Prediction neighborhoods are instead determined from geometric proximity in
% the aligned lattice.
Quantized current geometry is coded with G-PCC. Current and reference Gaussians
are stably Morton ordered only for deterministic serialization; temporal
neighborhoods are determined from geometric proximity in the aligned lattice,
not from Morton order.

\vspace{-5mm}
% \paragraph{Shared SH representation.}
% SH appearance accounts for most of the Gaussian attribute dimensionality and
% contains substantial cross-coefficient correlation. Asking the temporal
% predictor and residual coder to operate directly on all 48 coefficients would
% therefore require them to model both static coefficient redundancy and temporal
% variation. InterGS+ instead learns a fixed global PCA representation. For
% native SH vector $\mathbf c_{t,i}\in\mathbb R^{48}$,
\paragraph{Shared SH representation.}
SH appearance accounts for most of the Gaussian attribute dimensionality and
is strongly correlated. Rather than model both coefficient correlation and
temporal variation directly in the native 48-D space, InterGS+ uses a fixed
global PCA representation. For native SH vector
$\mathbf c_{t,i}\in\mathbb R^{48}$,
\vspace{-3mm}
\begin{equation}
\mathbf z_{t,i}
=
\mathbf U(\mathbf c_{t,i}-\boldsymbol{\mu})
\in\mathbb R^{38},
\qquad
\hat{\mathbf c}_{t,i}
=
\mathbf U^{\top}\hat{\mathbf z}_{t,i}
+\boldsymbol{\mu}.
\label{eq:pca}
\vspace{-3mm}
\end{equation}

% The fixed $48\!\rightarrow\!38$ transform is learned offline and shared by
% encoder and decoder.  We retain 38 PCA components based on validation: ranks
% 24 and 28 introduced a high-rate PSNR ceiling, whereas 38 removed this
% limitation while preserving approximately 99.99\% of the SH variance. The resulting compact, decorrelated representation is used for SH temporal prediction and SH residual coding, avoiding sequence-dependent transform signaling. Inverse PCA is applied only when an SH prediction or reconstruction is required in the native 48-D representation.
The fixed global $48\!\rightarrow\!38$ PCA transform is learned once offline
and reused unchanged across all sequences by both encoder and decoder, without
sequence-specific transform learning or signaling. We retain 38 PCA components
based on validation: ranks 24 and 28 introduced a high-rate PSNR ceiling,
whereas 38 removed this limitation while preserving approximately 99.99\% of
the SH variance. The resulting compact, decorrelated representation is used for
SH temporal prediction and SH residual coding. Inverse PCA is applied only when
an SH prediction or reconstruction is required in the native 48-D representation.

\vspace{-5mm}
\subsection{Hybrid Inter-Frame Prediction}
\label{subsec:prediction}
\vspace{-3mm}

% A single temporal estimator is unlikely to be optimal throughout a dynamic
% scene. When geometric correspondence is strong, one nearby reference Gaussian
% may already provide an accurate estimate; when correspondence is less certain,
% aggregation over several neighbors can be more robust. SH appearance is more
% challenging because its temporal evolution can depend jointly on reference
% appearance and neighborhood geometry. InterGS+ therefore retains the
% lightweight predictor bank of InterGS-Lite and introduces additional learned
% capacity only for SH.
% For current Gaussian $i$, reconstructed reference Gaussians are ranked by
A single temporal estimator is unlikely to be optimal throughout a dynamic
scene. When geometric correspondence is strong, a nearby reference Gaussian
may provide an accurate estimate; when it is less certain, aggregation over
several neighbors can be more robust. SH appearance is more challenging because
its temporal evolution depends jointly on reference appearance and neighborhood
geometry. InterGS+ therefore retains the lightweight InterGS-Lite predictor bank
and adds learned capacity only for SH. For current Gaussian $i$, reconstructed
reference Gaussians are ranked by $d_{ik}=\|\bar{\mathbf q}^{\,t}_{t-1,n_{ik}}-\mathbf q_{t,i}\|_2$. The classical branch uses the first $K_h=8$ reference Gaussians as eight
independent KNN candidates. A bilateral candidate aggregates the same
neighborhood using geometry-derived affinities:
% \vspace{-3mm}
% \[
% d_{ik}
% =
% \left\|
% \bar{\mathbf q}^{\,t}_{t-1,n_{ik}}
% -
% \mathbf q_{t,i}
% \right\|_2 .
% \]

\vspace{-4mm}
\begin{equation}
a_{ik}
=
\frac{\exp(-d_{ik}^{2}/2\sigma_q^{2})}
{\sum_{\ell=1}^{K_h}\exp(-d_{i\ell}^{2}/2\sigma_q^{2})},
\qquad
\tilde{\mathbf u}^{\,\mathrm{bil}}_{t,i}
=
\sum_{k=1}^{K_h}
a_{ik}\hat{\mathbf u}_{t-1,n_{ik}} .
\label{eq:bilateral}
\vspace{-3mm}
\end{equation}

% These predictors impose a strong locality prior at very low complexity, but
% their aggregation rule is fixed once geometric distances are known. This is
% restrictive for high-dimensional SH appearance, where prediction can depend on
% interactions among both reference content and local geometry.
These low-complexity predictors impose a fixed locality prior, which can be
restrictive for high-dimensional SH appearance where prediction depends jointly
on reference content and local geometry.

\vspace{-4mm}
\paragraph{Geometry-aware SH transformer.}
InterGS+ therefore augments the SH predictor bank with a lightweight
geometry-aware transformer operating in the PCA domain. We use $K_h=8$ for the classical predictors to retain a compact local search, while the transformer uses $K_T=32$ reconstructed reference neighbors to provide broader context for modeling geometry-aware SH interactions. Each token combines the 38-D reference SH-PCA representation with its 3-D offset from the current Gaussian
and is gated by geometric affinity:
% InterGS+ augments the SH predictor bank with a lightweight transformer operating
% in the PCA domain. The classical branch uses $K_h=8$ neighbors for compact local
% search, whereas the transformer uses $K_T=32$ to provide broader context for
% geometry-aware SH interactions. Each token concatenates the 38-D reference
% SH-PCA representation with its 3-D offset from the current Gaussian and is
% gated by geometric affinity:
\vspace{-4mm}
\begin{equation}
\Delta\mathbf q_{ik}
=
\bar{\mathbf q}^{\,t}_{t-1,n_{ik}}
-
\mathbf q_{t,i},
\qquad
\mathbf t_{ik}
=
[\hat{\mathbf z}_{t-1,n_{ik}};\Delta\mathbf q_{ik}]
\in\mathbb R^{41},
\qquad
\tilde{\mathbf t}_{ik}=a_{ik}\mathbf t_{ik}.
\label{eq:transformer_token}
\vspace{-3mm}
\end{equation}

Affinity gating provides an explicit geometric-locality prior before
self-attention. Content attention alone, however, does not identify the
geometric role of each neighbor. Hence, PE-Block injects geometry at
both the token and attention-logit levels:
% Affinity gating provides a locality prior, while the PE-Block injects geometry
% at both the token and attention-logit levels:
\vspace{-4mm}
\begin{equation}
\mathbf p_{ik}
=
\boldsymbol{\rho}_k
+
\phi_{\mathrm{geo}}(\Delta\mathbf q_{ik})
+
\phi_F(\Delta\mathbf q_{ik}),
\qquad
s_{irk}
=
\frac{\mathbf q_{ir}^{\top}\mathbf k_{ik}}{\sqrt{E}}
+
\phi_{\mathrm{bias}}
(\Delta\mathbf q_{ik}-\Delta\mathbf q_{ir}),
\quad E=64 .
\label{eq:peblock}
\end{equation}

% Here, $\boldsymbol{\rho}_k$ is a learned neighbor-rank embedding,
% $\phi_{\mathrm{geo}}$ embeds the signed geometry offset, and $\phi_F$ denotes
% the additive multi-frequency Fourier encoding. The pairwise term modifies the
% attention logits according to relative geometry, allowing interactions among
% reference Gaussians to depend explicitly on their spatial configuration.

% The model uses single-head attention followed by residual normalization and a
% $64\!\rightarrow\!128\!\rightarrow\!64$ GeLU feed-forward block. Rather than
% symmetrically pooling all output tokens, the refined first-ranked token is
% mapped to the predicted 38-D PCA SH vector. The nearest reference Gaussian
% therefore remains the local anchor while incorporating information from the
% complete 32-neighbor context. The model contains 53,964 trainable parameters
% and is trained offline with an $\ell_1$ prediction loss using temporal offsets
% of one, two, and three frames weighted by $0.25$, $0.5$, and $0.25$,
% respectively. Once trained, its parameters are fixed and shared by encoder and
% decoder. Overall, the predictor maps a $32\times41$ neighborhood representation
% to $32\times64$ latent tokens, applies geometry-aware self-attention and the
% feed-forward block, and projects the refined first-ranked token to the 38-D PCA
% SH prediction.
Here, $\rho_k$ is a learned neighbor-rank embedding, $\phi_{\rm geo}$ embeds
the signed geometry offset, $\phi_F$ provides additive multi-frequency Fourier
features, and $\phi_{\rm bias}$ introduces pairwise relative-position bias in
the attention logits, allowing interactions to depend explicitly on spatial
configuration. The model uses single-head attention followed by residual
normalization and a $64\!\rightarrow\!128\!\rightarrow\!64$ GeLU feed-forward
block. The refined first-ranked token is projected to the 38-D PCA SH
prediction, preserving the nearest Gaussian as the local anchor while
incorporating the full 32-neighbor context. The model contains 53,964 trainable
parameters and is trained offline with an $\ell_1$ prediction loss using
temporal offsets $\{1,2,3\}$ weighted by $\{0.25,0.5,0.25\}$. Once trained,
its parameters are fixed and shared by encoder and decoder. Overall, the
predictor maps a $32\times41$ neighborhood representation to $32\times64$
latent tokens and outputs a 38-D PCA SH prediction.
\begin{figure}[t]
    \centering
    \begin{minipage}[c]{0.60\linewidth}
        \centering
        \includegraphics[
            width=\linewidth,
            trim=12 12 12 12,
            clip
        ]{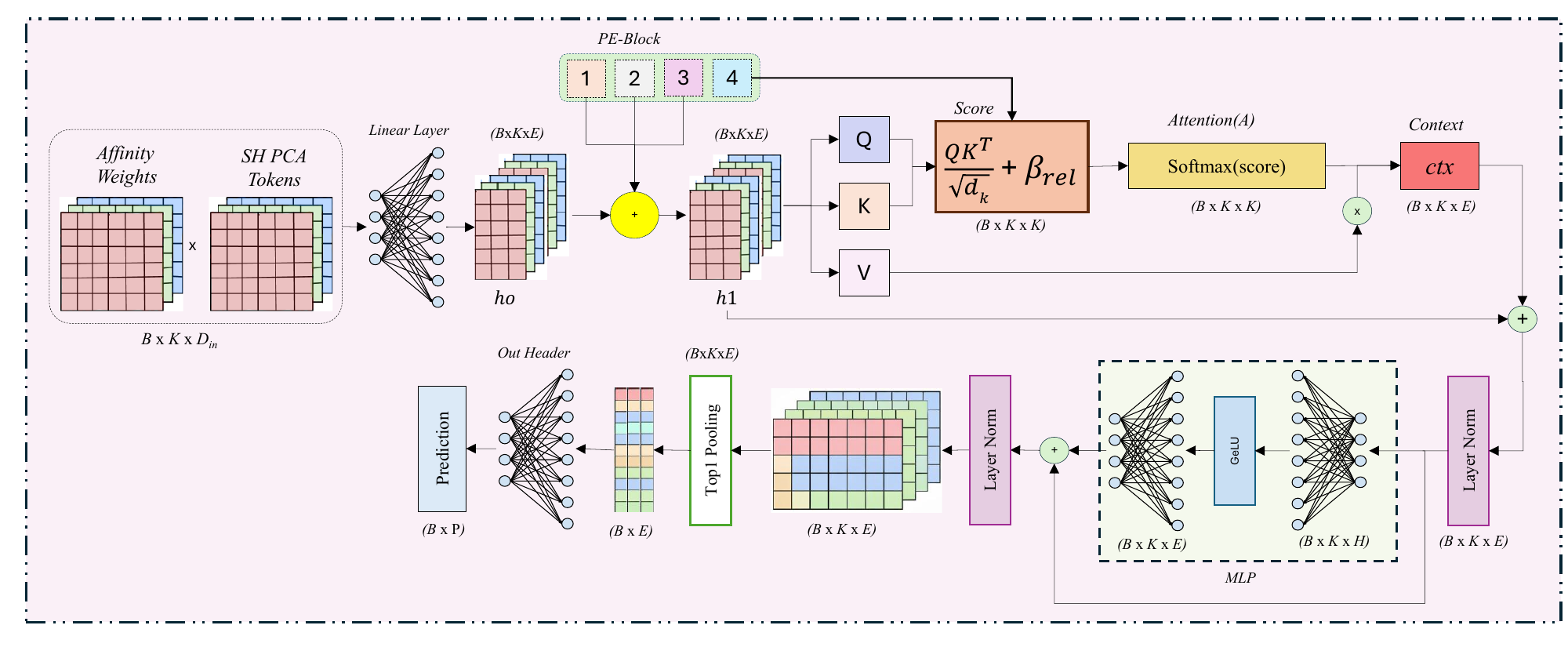}
    \end{minipage}
    \hfill
    \begin{minipage}[c]{0.36\linewidth}
        \centering
        \includegraphics[
            width=\linewidth,
            trim=12 12 12 12,
            clip
        ]{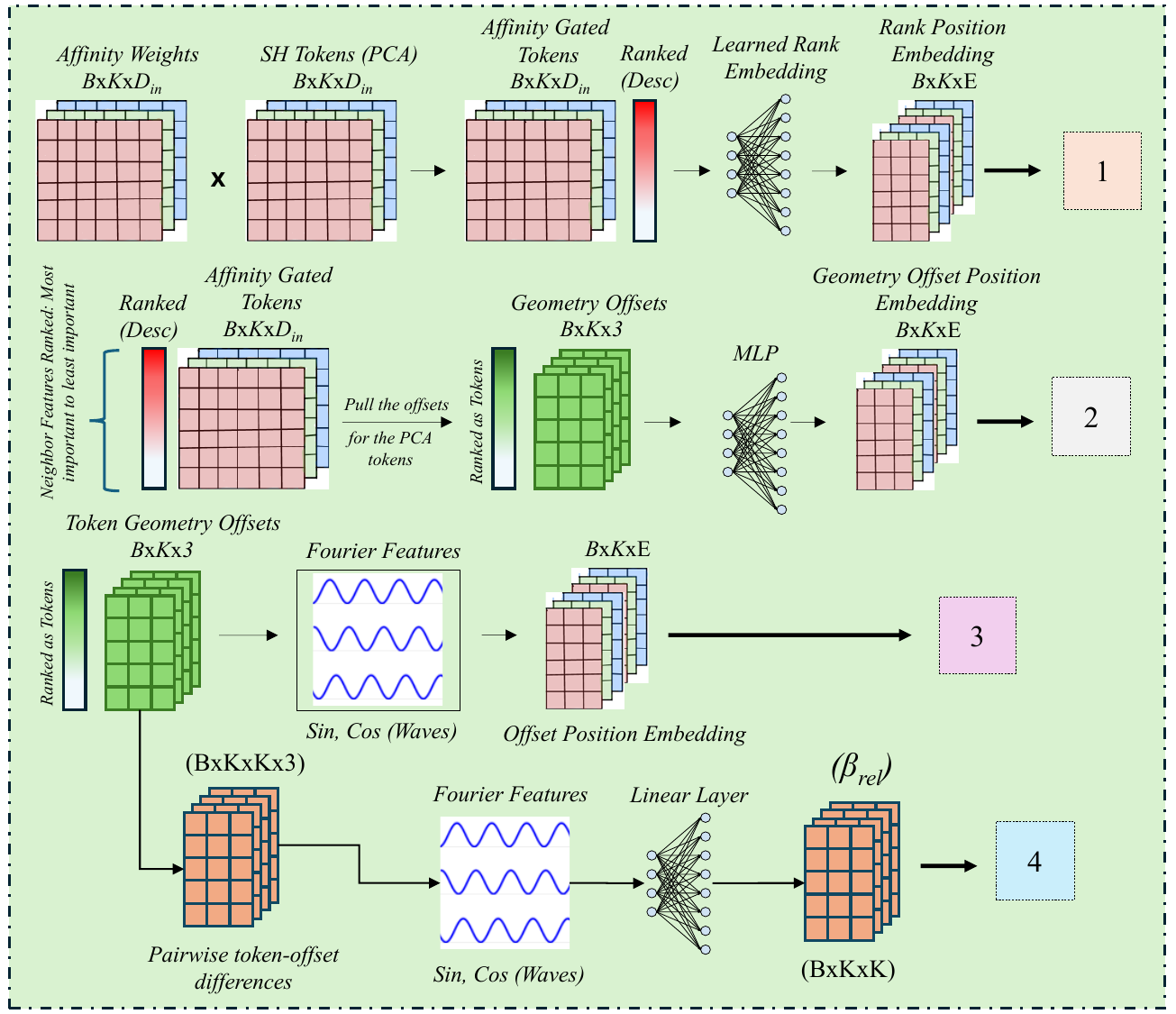}
    \end{minipage}

    \vspace{-2mm}
    \caption{Geometry-aware SH predictor in InterGS+:
    (a) lightweight transformer and (b) PE-Block incorporating neighbor rank,
    geometry-offset embedding, additive Fourier encoding, and relative-position
    bias.}
    \label{fig:transformer_pe}
    \vspace{-5mm}
\end{figure}
\vspace{-5mm}
\paragraph{Adaptive predictor selection.}
% Additional model capacity is useful only where it reduces temporal prediction
% error. SH therefore selects among the bilateral candidate, eight KNN
% candidates, and the transformer, whereas non-SH attributes select among the
% bilateral and eight KNN candidates. Transformer predictions are inverse-PCA
% mapped when native-domain SH error is required for mode comparison; after
% selection, the chosen SH prediction is represented in the shared PCA domain for
% residual coding. The SH and non-SH decisions are packed into one control symbol:
SH selects among the bilateral candidate, eight KNN candidates, and the
transformer, whereas non-SH attributes select among the bilateral and eight KNN
candidates. Transformer predictions are inverse-PCA mapped only when
native-domain SH error is required for mode comparison; after selection, the
chosen SH prediction is represented in the shared PCA domain for residual
coding. The SH and non-SH decisions are packed into one control symbol:
\vspace{-3mm}
\begin{equation}
p_{t,i}
=
m^{\mathrm{SH}}_{t,i}(K_h+1)
+
m^{\mathrm{NS}}_{t,i},
\qquad K_h=8.
\label{eq:mode_pack}
\vspace{-3mm}
\end{equation}

% The ten SH modes and nine non-SH modes yield 90 possible packed symbols,
% corresponding to a fixed-length upper bound of 7 bits/Gaussian before lossless
% compression. Only the selected mode is transmitted; the corresponding
% prediction is regenerated from the reconstructed temporal reference at the
% decoder.
The 10 SH and 9 non-SH modes yield 90 possible symbols, corresponding to a
fixed-length upper bound of 7 bits/Gaussian before lossless compression. Only
the selected mode is transmitted; the decoder regenerates its prediction from
the reconstructed temporal reference.

\vspace{-4mm}
\subsection{Attribute-Adaptive Residual Coding}
\label{subsec:residual}
\vspace{-3mm}

Temporal prediction converts Gaussian attributes into conditional innovation
signals, but these residuals do not share the same statistics or
parameterization constraints. InterGS+ therefore avoids imposing a common
residual representation and instead matches residual formation and coding to
the structure of each attribute.

\vspace{-5mm}
\paragraph{SH: PCA-domain VQ with refinement.}
For the selected SH predictor, both the current appearance and its temporal
prediction are represented in the shared PCA domain. The SH innovation is
therefore
% \vspace{-3mm}
% \begin{equation}
$
\mathbf r^{\mathrm{SH}}_{t,i}
=
\mathbf z_{t,i}
-
\tilde{\mathbf z}_{t,i}
\in\mathbb R^{38}.
\label{eq:sh_pca_residual}$ The first three and remaining 35 PCA components are coded as separate vector
groups. Because this partition is applied after the joint $48\!\rightarrow\!38$
transform, these groups denote PCA-component groups rather than the original
native SH DC and AC coefficients. For each group $g$, fixed global
normalization statistics associated with the shared VQ codebook are applied as
$\mathbf y_i^{(g)}=(\mathbf r_i^{(g)}-\boldsymbol{\mu}_g)/
\boldsymbol{\sigma}_g$, where $\boldsymbol{\mu}_g$ and
$\boldsymbol{\sigma}_g$ are shared by encoder and decoder and are therefore not
transmitted per frame. The normalized residual is then assigned to its nearest
K-means centroid. The 3-D and 35-D groups use 42 and 8 centroids,
respectively. A finite codebook captures recurring multidimensional residual
structure, while a scalar refinement layer preserves the remaining deviations:
% \vspace{-3mm}
% \end{equation}

% The first three and remaining 35 PCA components are coded as separate vector
% groups. Because this partition is applied after the joint $48\!\rightarrow\!38$
% transform, these groups denote PCA-component groups rather than the original
% native SH DC and AC coefficients. For each group $g$, fixed global statistics associated with the shared VQ codebook normalize the residual vectors,
% \vspace{-3mm}
% \[
% \vspace{-3mm}
% \mathbf y_i^{(g)}
% =
% \frac{\mathbf r_i^{(g)}-\boldsymbol{\mu}_g}
% {\boldsymbol{\sigma}_g},
% \]
% after which the nearest K-means centroid is selected. The 3-D and 35-D groups
% use 42 and 8 centroids, respectively. 

% The first three and remaining 35 PCA components are coded as separate vector
% groups. Because this partition is applied after the joint $48\!\rightarrow\!38$
% transform, these groups denote PCA-component groups rather than the original
% native SH DC and AC coefficients. For each group $g$, fixed global
% normalization statistics associated with the shared VQ codebook are applied:
% \vspace{-3mm}
% \[
% \mathbf y_i^{(g)}
% =
% \frac{\mathbf r_i^{(g)}-\boldsymbol{\mu}_g}
% {\boldsymbol{\sigma}_g},
% \]
% where $\boldsymbol{\mu}_g$ and $\boldsymbol{\sigma}_g$ are shared by encoder
% and decoder and are therefore not transmitted per frame. The normalized
% residual is then assigned to its nearest K-means centroid. The 3-D and 35-D
% groups use 42 and 8 centroids, respectively. A finite codebook captures recurring multidimensional residual structure, while a scalar refinement layer preserves the remaining deviations:
\vspace{-7mm}
\begin{equation}
k_i^{(g)}
=
\arg\min_k
\left\|
\mathbf y_i^{(g)}
-
\mathbf c_k^{(g)}
\right\|_2^2,
\qquad
\mathbf l_i^{(g)}
=
\operatorname{round}
\left(
\frac{
\mathbf r_i^{(g)}-\bar{\mathbf r}_i^{(g)}
}{\delta_g}
\right),
\quad g\in\{1,2\}.
\label{eq:sh_vq}
\vspace{-3mm}
\end{equation}

The decoded VQ vector and refinement reconstruct the 38-D PCA innovation,
which is added to the selected PCA-domain prediction. Inverse PCA is then
applied to obtain the reconstructed 48-D SH coefficients. This preserves the
dimensionality reduction throughout both temporal prediction and residual
coding rather than using PCA only as an intermediate predictor representation.

\vspace{-5mm}
\paragraph{Opacity and scale: activated-domain companding.}
Opacity and scale require a different residual domain because their stored
parameters are transformed before rendering: opacity through a sigmoid and
scale through an exponential. Equal perturbations in stored parameter space
therefore do not correspond to equal changes in the rendered Gaussian. Let
$A_o(o)=\sigma(o)$,
$A_s(s)=\exp(s)$, and
$C_{\gamma}(x)=x^{\gamma}$.
For $g\in\{o,s\}$, InterGS+ forms
\vspace{-6mm}
\begin{equation}
\mathbf r^{g}_{t,i}
=
C_{\gamma_g}
\!\left(A_g(\mathbf f^g_{t,i})\right)
-
C_{\gamma_g}
\!\left(A_g(\tilde{\mathbf f}^g_{t,i})\right),
\qquad
\mathbf l^{g}_{t,i}
=
\operatorname{round}
\left(
\frac{\mathbf r^g_{t,i}}{\delta_g}
\right).
\label{eq:activated_residual}
\vspace{-5mm}
\end{equation}

The power transform changes local quantization sensitivity according to
$d x^\gamma/dx=\gamma x^{\gamma-1}$. Importantly, companding is applied
independently to the current and predicted attributes before subtraction;
applying it to an already formed raw residual would define a different coding
domain. Reconstruction therefore adds the decoded innovation in the companded
domain before applying the inverse power transform and inverse activation.

\vspace{-5mm}
\paragraph{Rotation: canonicalization, VQ, and refinement.}
Rotation cannot be treated as an unconstrained four-dimensional signal.
Quaternions $\mathbf q$ and $-\mathbf q$ represent the same physical
orientation, while scale and rotation jointly determine the anisotropic
Gaussian covariance, allowing equivalent parameter tuples to represent the
same Gaussian. InterGS+ first applies deterministic joint scale--rotation canonicalization.
Current and predicted quaternions are then normalized, after which the current
quaternion is sign-aligned with its prediction:
\vspace{-3mm}
\begin{equation}
\mathbf q_{t,i}
\leftarrow
\begin{cases}
-\mathbf q_{t,i},
&
\mathbf q_{t,i}^{\top}\tilde{\mathbf q}_{t,i}<0,
\\
\mathbf q_{t,i},
&
\text{otherwise},
\end{cases}
\qquad
\mathbf r^q_{t,i}
=
\mathbf q_{t,i}
-
\tilde{\mathbf q}_{t,i}.
\label{eq:rotation_residual}
\vspace{-3mm}
\end{equation}

This removes parameter differences that do not correspond to changes in
physical orientation. The resulting four-dimensional innovation is coded
using a per-frame 32-centroid vector quantizer followed by uniform scalar
refinement of the remaining VQ error. After residual reconstruction, the
innovation is added to the predicted quaternion and the result is renormalized
to unit length.

The three residual branches therefore follow the same principle: temporal
prediction removes information that can be inferred from the reconstructed
reference, while the remaining innovation is represented in a domain matched
to the statistics, rendering behavior, or structural constraints of the
corresponding Gaussian attribute.

\vspace{-5mm}
\paragraph{Bitstream and predictive reconstruction.}
% Each predictive frame contains the G-PCC geometry stream, packed
% predictor-control symbols, SH PCA-group VQ indices and refinement data,
% opacity and scale residual symbols, rotation VQ and refinement data, and the
% side information required to invert the corresponding transforms. The global
% PCA basis and transformer weights are fixed and shared and are therefore not
% transmitted per frame. Predictor symbols, VQ indices, refinement symbols, and
% attribute side information are losslessly compressed using the 7z/LZMA
% backend, while geometry remains a separate G-PCC stream.

% At the decoder, temporal neighborhoods and predictor candidates are reproduced
% from the reconstructed reference. The transmitted mode selects the appropriate
% prediction, decoded residual information reconstructs the current attributes,
% and inverse PCA recovers native SH coefficients after PCA-domain SH
% reconstruction. The resulting Gaussian frame then becomes the reference for
% the next predictive frame.
Each P-frame transmits G-PCC geometry, packed predictor modes, SH PCA-domain
VQ/refinement data, non-SH residual streams, and the required side information;
the fixed PCA basis and transformer weights are shared and not transmitted.
Attribute and control streams are losslessly compressed with 7z/LZMA. The
decoder regenerates the selected prediction from the reconstructed reference,
decodes the residuals, applies inverse PCA to recover SH, and uses the resulting
frame as the next temporal reference.

\vspace{-6mm}
\section{Experimental Results}
\label{sec:experiments}
\vspace{-5mm}

\subsection{Experimental Setup}
\label{subsec:setup}
\vspace{-3mm}

We evaluate InterGS+ on \textbf{10 dynamic 3DGS sequences}, coding six
consecutive frames (0--5) from each sequence. The test set contains the
semi-tracked \emph{Bartender}, \emph{Cinema}, and \emph{Breakfast} sequences,
their tracked counterparts, together with \emph{ManWithFruit-tracked},
\emph{Trio}, \emph{Makeup}, and \emph{Musical}. InterGS+ is compared with
\textbf{G-PCC Amd.~1} under octree-raht (intra coding)~\cite{MPEG_GSC_CTC} and
\textbf{InterGS-Lite}~\cite{Talha2026InterGSLite}; the original InterGS is omitted
because InterGS-Lite already improves upon it within the same post-hoc
inter-frame framework. All comparisons follow the MPEG CTC for the dataset,
rendering views, distortion metric, and evaluation conditions; results are
reported over six consecutive frames (0--5) for all three methods.

\vspace{-5mm}
\paragraph{Training and test split.}
The SH predictor is trained only on semi-tracked \emph{Bartender},
\emph{Cinema}, and \emph{Breakfast}, using frames 15--30. Evaluation on these
sequence identities uses the disjoint range 0--5, while the remaining seven
test sequences are unseen during training. Training uses
$\Delta t\in\{1,2,3\}$ with weights $0.25$, $0.5$, and $0.25$ under an
$L_1$ loss in the global SH PCA domain. The trained model is fixed for all
test sequences. All model parameters and coding hyperparameters are fixed globally and applied
unchanged to every evaluated sequence; no sequence-dependent parameter tuning
or manual content-specific adjustment is performed. Unless otherwise stated, InterGS+ uses 18-bit geometry,
the fixed $48\!\rightarrow\!38$ SH PCA transform, $K_h=8$ classical
neighbors, and $K_T=32$ transformer neighbors. Bitrate includes geometry,
residuals, predictor signaling, codebooks, and side information; reconstruction
quality is measured using PSNR and summarized by BD-rate and BD-PSNR.

\begin{figure}[t]
    \centering

    % ---------------- Row 1 ----------------
    \begin{minipage}[t]{0.19\linewidth}
        \centering
        \includegraphics[width=\linewidth]
        {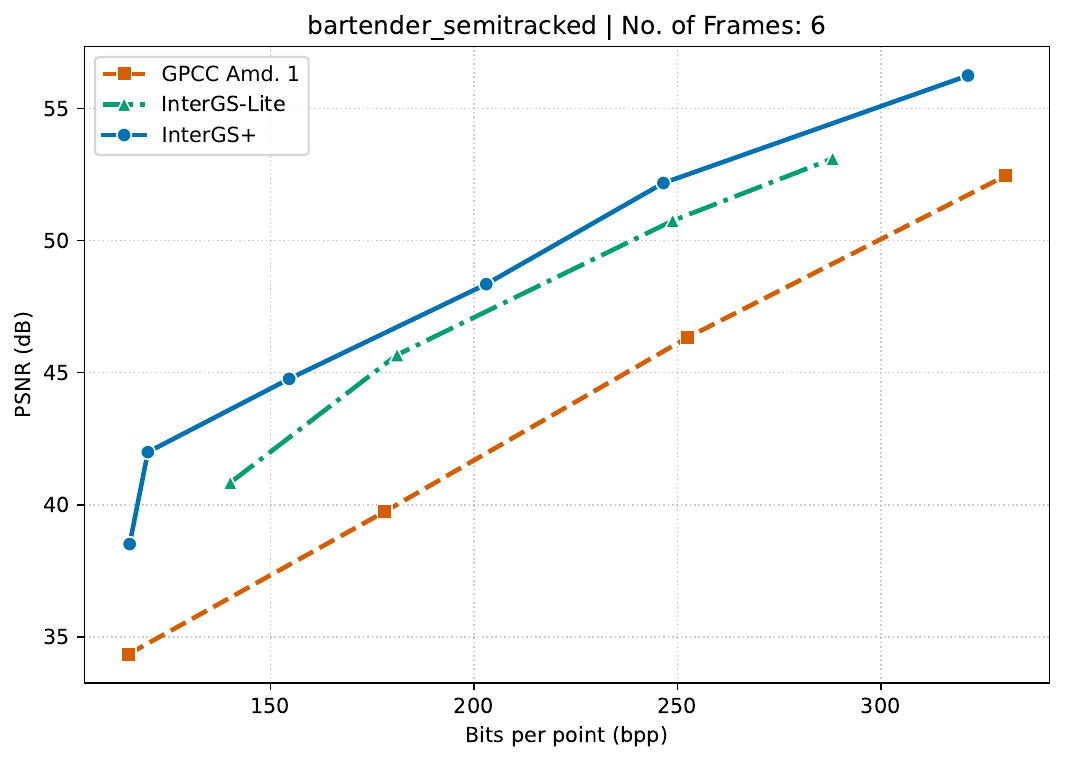}
    \end{minipage}
    \hfill
    \begin{minipage}[t]{0.19\linewidth}
        \centering
        \includegraphics[width=\linewidth]
        {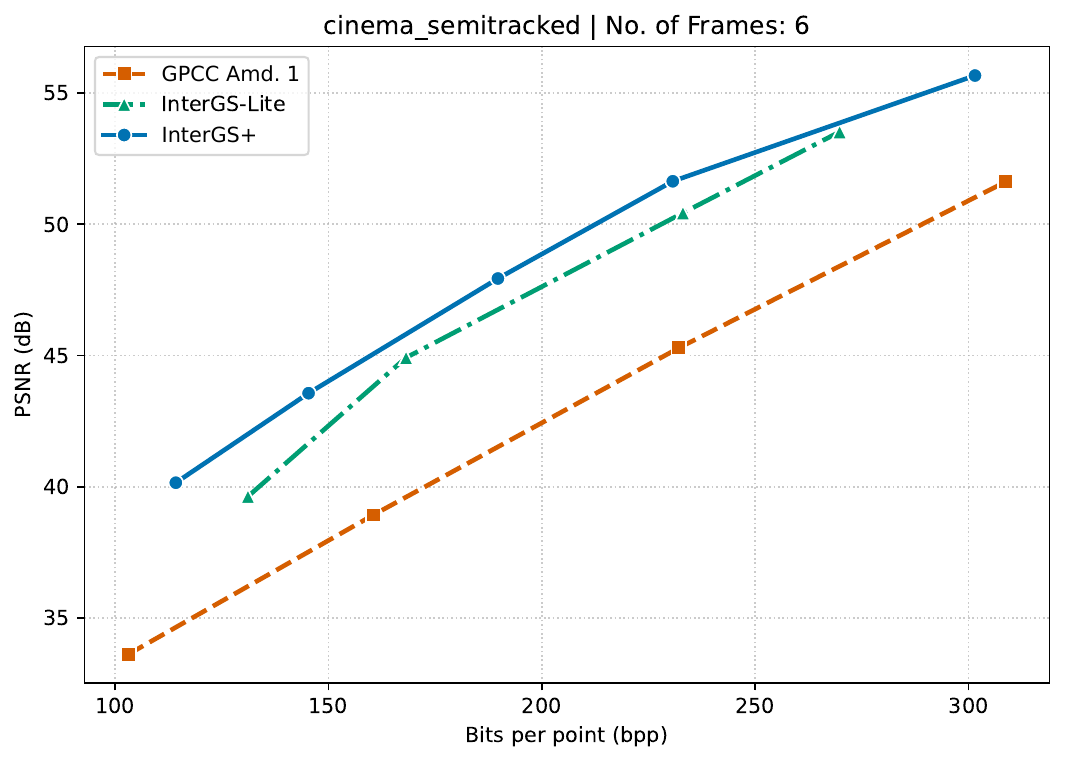}
    \end{minipage}
    \hfill
    \begin{minipage}[t]{0.19\linewidth}
        \centering
        \includegraphics[width=\linewidth]
        {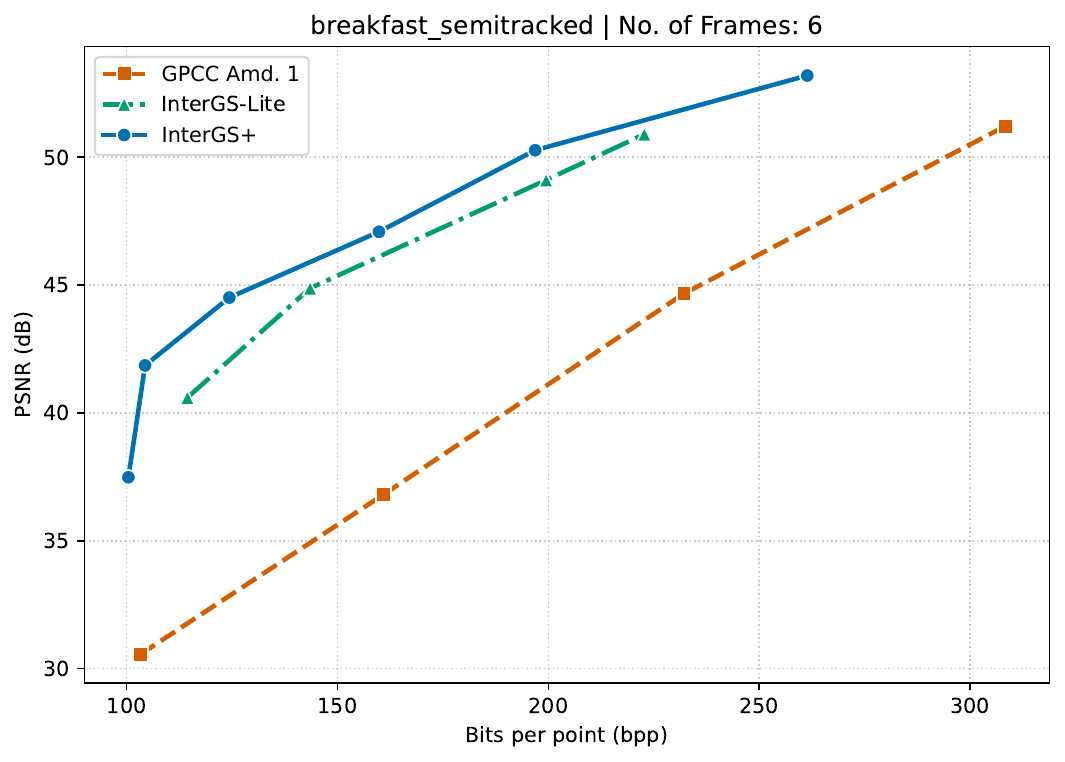}
    \end{minipage}
    \hfill
    \begin{minipage}[t]{0.19\linewidth}
        \centering
        \includegraphics[width=\linewidth]
        {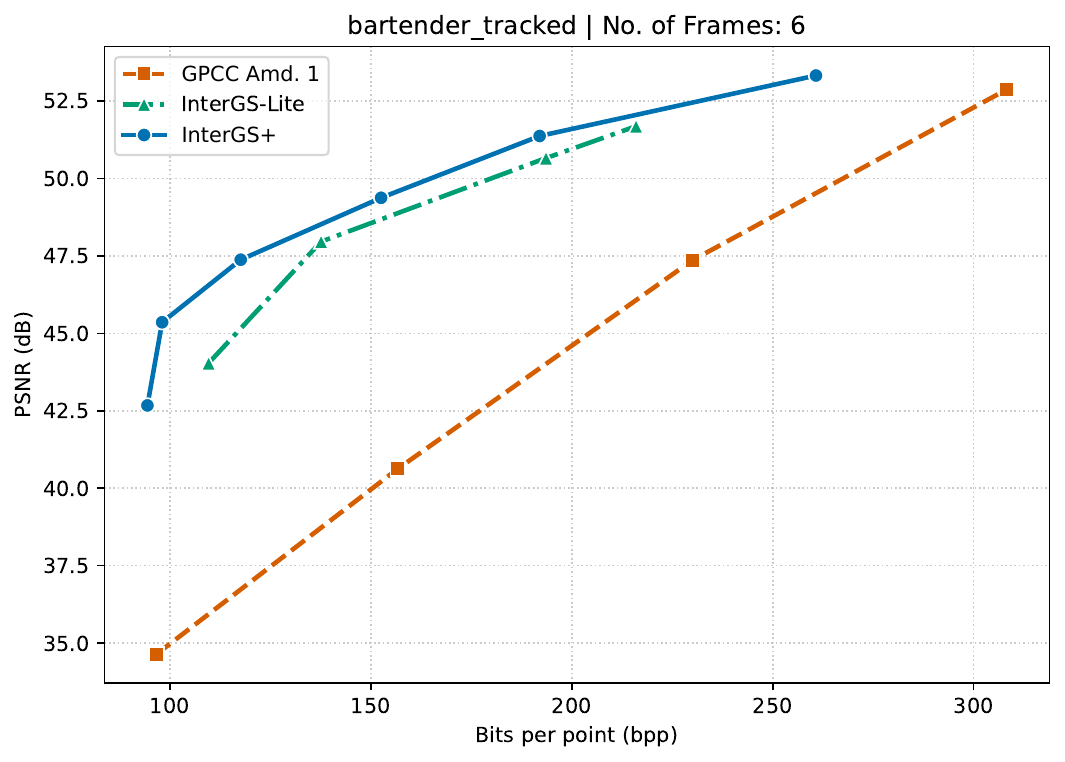}
    \end{minipage}
    \hfill
    \begin{minipage}[t]{0.19\linewidth}
        \centering
        \includegraphics[width=\linewidth]
        {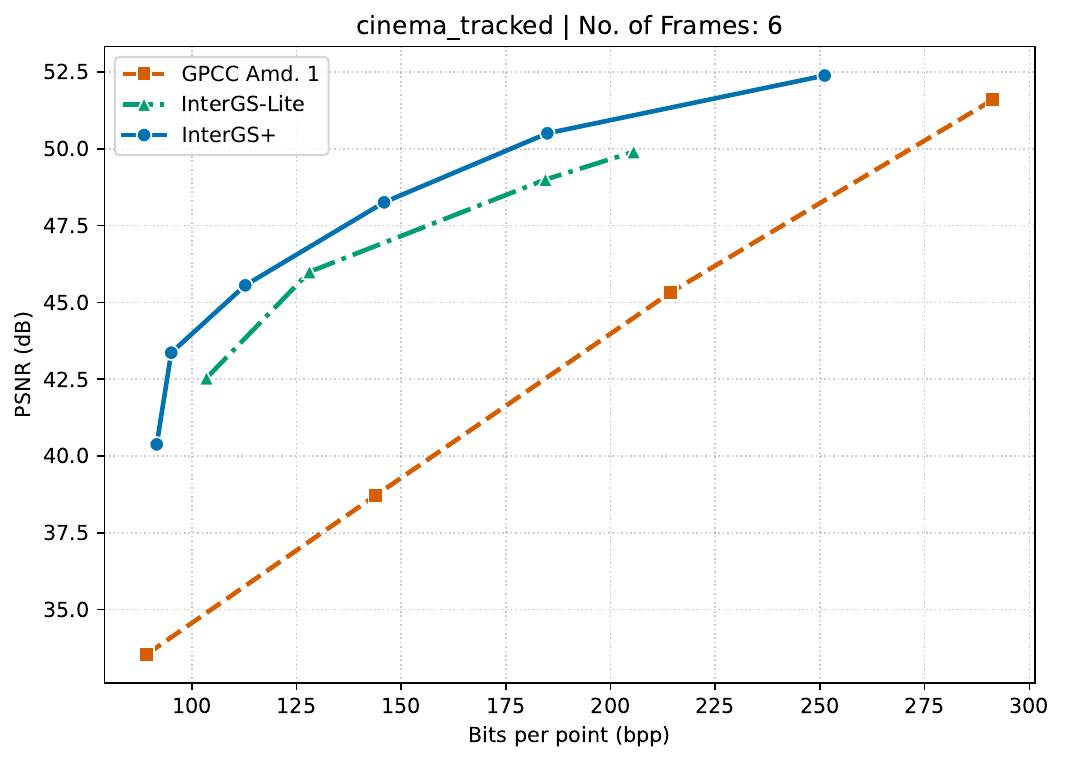}
    \end{minipage}

    \vspace{1mm}

    % ---------------- Row 2 ----------------
    \begin{minipage}[t]{0.19\linewidth}
        \centering
        \includegraphics[width=\linewidth]
        {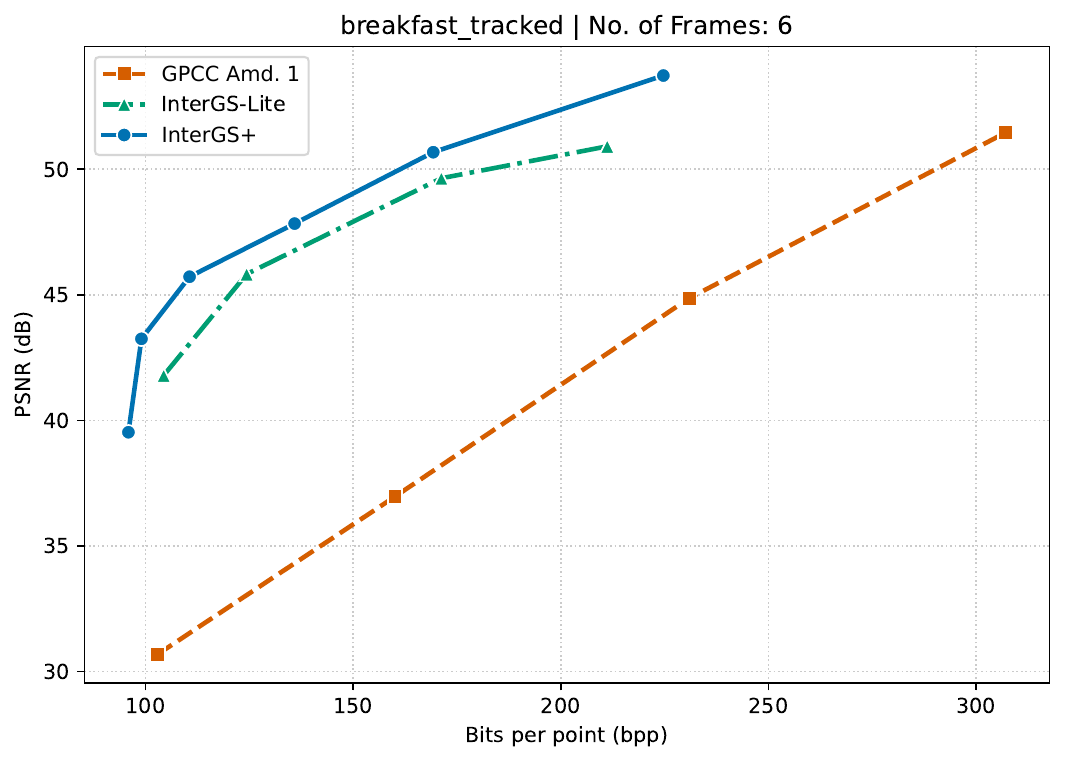}
    \end{minipage}
    \hfill
    \begin{minipage}[t]{0.19\linewidth}
        \centering
        \includegraphics[width=\linewidth]
        {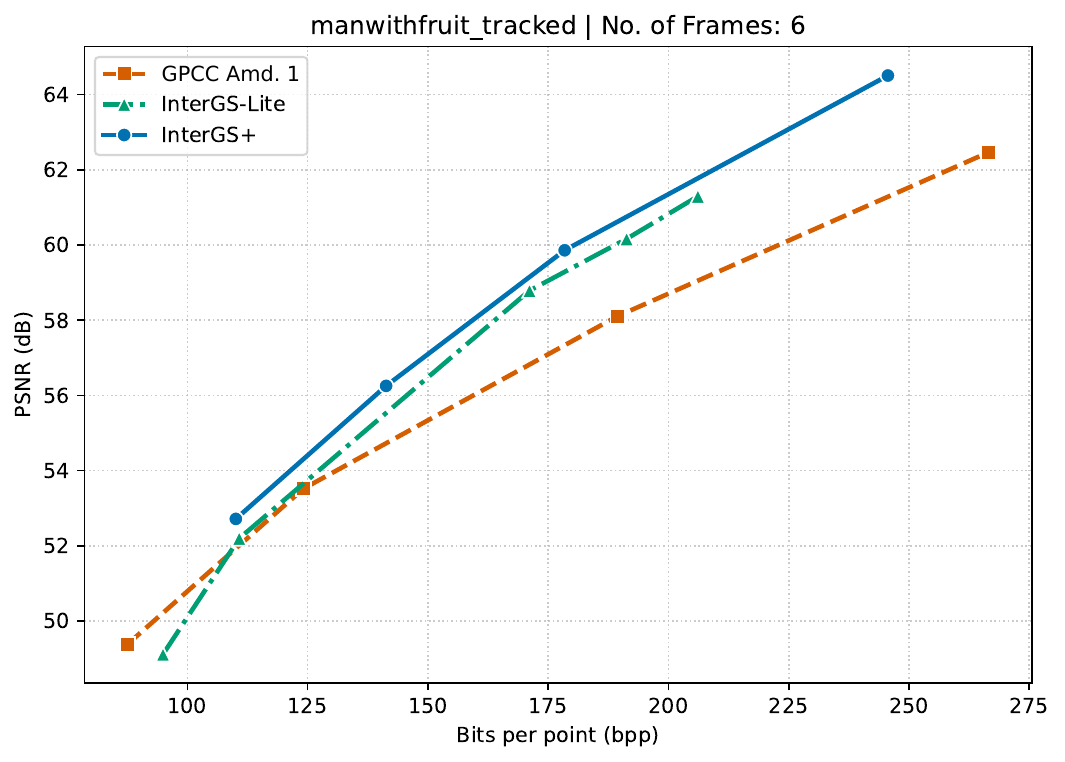}
    \end{minipage}
    \hfill
    \begin{minipage}[t]{0.19\linewidth}
        \centering
        \includegraphics[width=\linewidth]
        {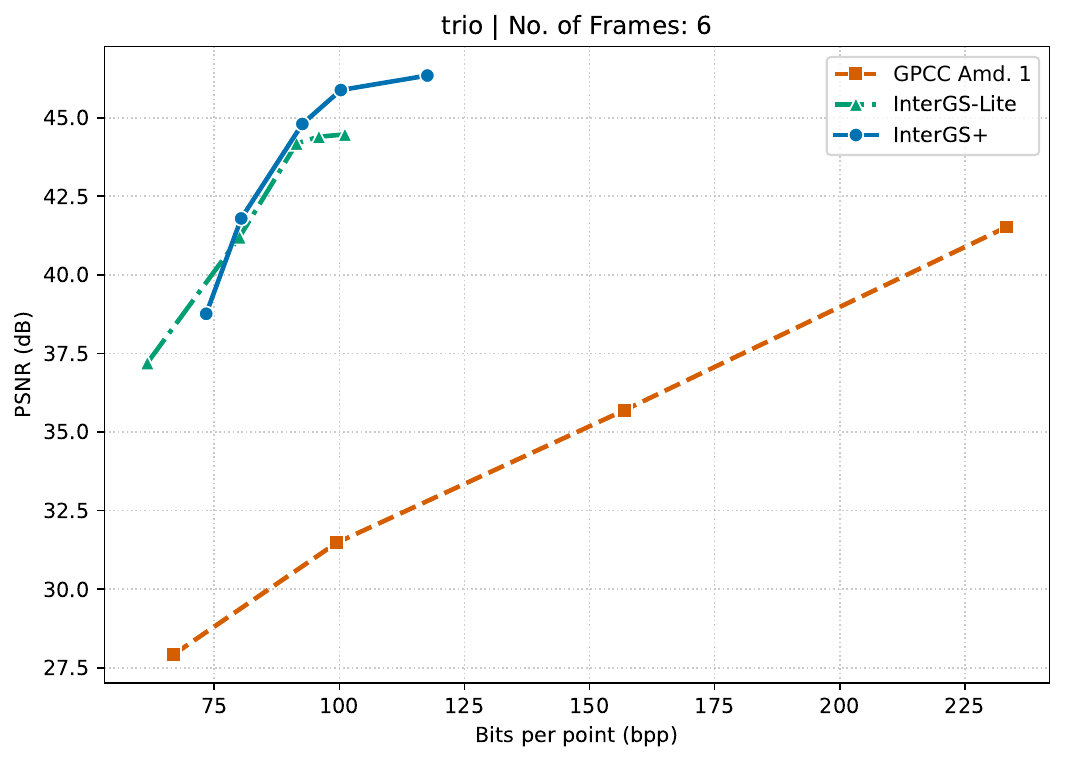}
    \end{minipage}
    \hfill
    \begin{minipage}[t]{0.19\linewidth}
        \centering
        \includegraphics[width=\linewidth]
        {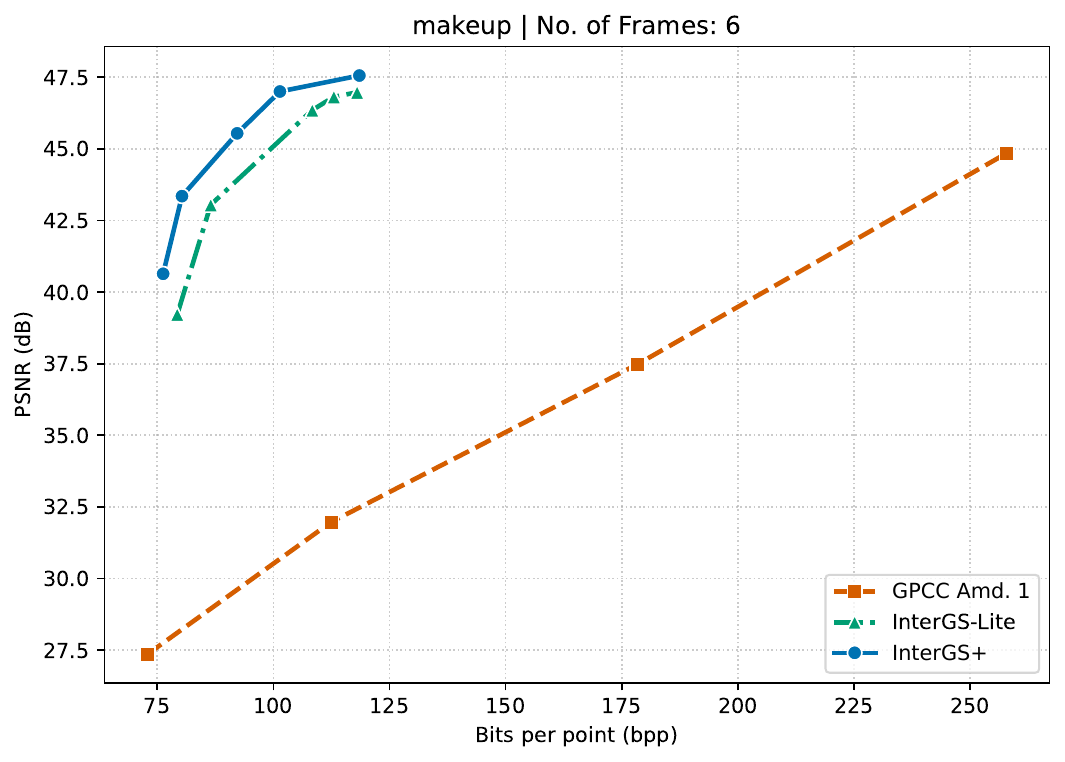}
    \end{minipage}
    \hfill
    \begin{minipage}[t]{0.19\linewidth}
        \centering
        \includegraphics[width=\linewidth]
        {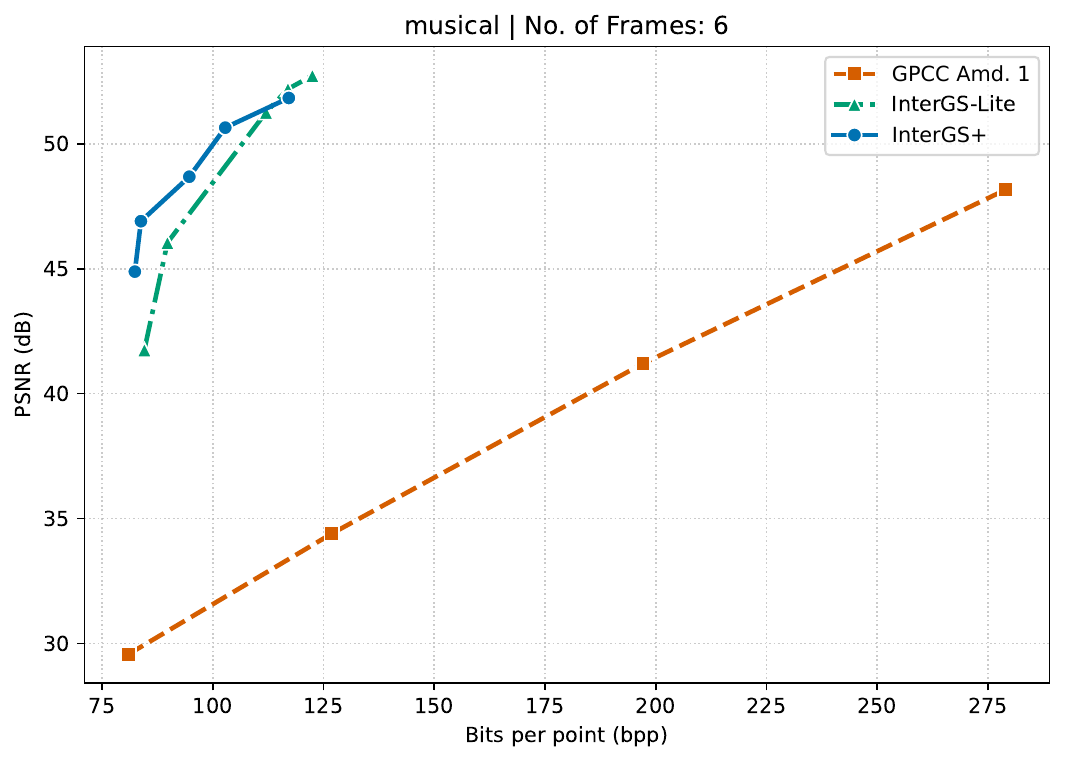}
    \end{minipage}

    \vspace{-4mm}
    \caption{Rate--distortion performance of G-PCC Amd.~1, InterGS-Lite,
    and InterGS+ on the 10 evaluated sequences.}
    \label{fig:rd_curves}
    \vspace{-3mm}
\end{figure}

\vspace{-5mm}
\paragraph{Rate--Distortion Performance.}
% Figure~\ref{fig:rd_curves} compares InterGS+ with G-PCC Amd.~1 and
% InterGS-Lite on all 10 sequences. InterGS+ consistently improves the
% rate--distortion tradeoff over G-PCC Amd.~1 and provides additional gains over
% InterGS-Lite across the evaluated content. Relative to G-PCC Amd.~1,
% InterGS+ achieves an average 45.48\% BD-rate reduction and
% 9.98 dB BD-PSNR gain. Improvements are observed on every sequence,
% including the seven sequences unseen during training. The largest gains over
% G-PCC Amd.~1 occur on the unseen \emph{Musical}, \emph{Makeup}, and
% \emph{Trio} sequences, with BD-rate reductions of $67.08\%$, $66.59\%$, and
% $64.38\%$, respectively. Even the object-centric
% \emph{ManWithFruit-tracked} sequence improves by $14.05\%$ BD-rate and
% $1.97$ dB BD-PSNR. Compared with InterGS-Lite, InterGS+ also provides consistent additional
% coding gains across the evaluated sequences. This indicates that the proposed
% PCA-domain geometry-aware SH prediction and attribute-specific residual coding
% provide benefits beyond the lightweight temporal prediction framework inherited
% from InterGS-Lite.
Figure~\ref{fig:rd_curves} compares InterGS+ with G-PCC Amd.~1 and
InterGS-Lite on all 10 sequences. InterGS+ improves the rate--distortion
tradeoff on every sequence, achieving average gains of 45.48\% BD-rate and
9.98 dB BD-PSNR over G-PCC Amd.~1, and 7.78\% BD-rate and 1.10 dB
BD-PSNR over InterGS-Lite. Gains also extend to all seven sequences unseen
during training; the largest reductions relative to G-PCC Amd.~1 occur on
\emph{Musical}, \emph{Makeup}, and \emph{Trio} at 67.08\%, 66.59\%, and
64.38\%, respectively, while \emph{ManWithFruit-tracked} still improves by
14.05\% BD-rate and 1.97 dB BD-PSNR. These results show that the
PCA-domain geometry-aware SH predictor and attribute-specific residual coding
provide consistent gains on top of InterGS-Lite temporal framework and
generalize beyond training content.

% ============================================================
% Combined BD metrics + predictor analysis
% ============================================================

\begin{figure}[t]
    \centering

    \begin{minipage}[c]{0.40\linewidth}
        \centering
        \scriptsize
        \setlength{\tabcolsep}{2.0pt}
        \renewcommand{\arraystretch}{0.88}

        \begin{tabular}{l|rr|rr}
        \hline
        & \multicolumn{2}{c|}{vs. G-PCC Amd.~1}
        & \multicolumn{2}{c}{vs. InterGS-Lite} \\
        \cline{2-3}\cline{4-5}
        Sequence & BD-R & BD-P & BD-R & BD-P \\
                 & (\%) & (dB) & (\%) & (dB) \\
        \hline
        Bartender-ST   & 32.20 & 6.34  & 10.10 & 1.47 \\
        Cinema-ST      & 29.74 & 6.01  & 7.06  & 1.23 \\
        Breakfast-ST   & 42.55 & 9.79  & 9.58  & 1.39 \\
        Bartender-T    & 43.99 & 8.30  & 9.21  & 0.95 \\
        Cinema-T       & 43.93 & 8.22  & 11.11 & 1.32 \\
        Breakfast-T    & 50.24 & 12.71 & 9.51  & 1.27 \\
        ManWithFruit-T & 14.05 & 1.97  & 2.61  & 0.37 \\
        Trio           & 64.38 & 13.23 & 2.45  & 0.53 \\
        Makeup         & 66.59 & 15.33 & 8.05  & 1.27 \\
        Musical        & 67.08 & 17.92 & 8.10  & 1.21 \\
        \hline
        Average &
        \textbf{45.48} & \textbf{9.98} &
        \textbf{7.78} & \textbf{1.10} \\
        \hline
        \end{tabular}

        \vspace{1mm}
        {\footnotesize (a) BD metrics}
    \end{minipage}
    \hfill
    \begin{minipage}[c]{0.52\linewidth}
        \centering
        \includegraphics[width=\linewidth]
        {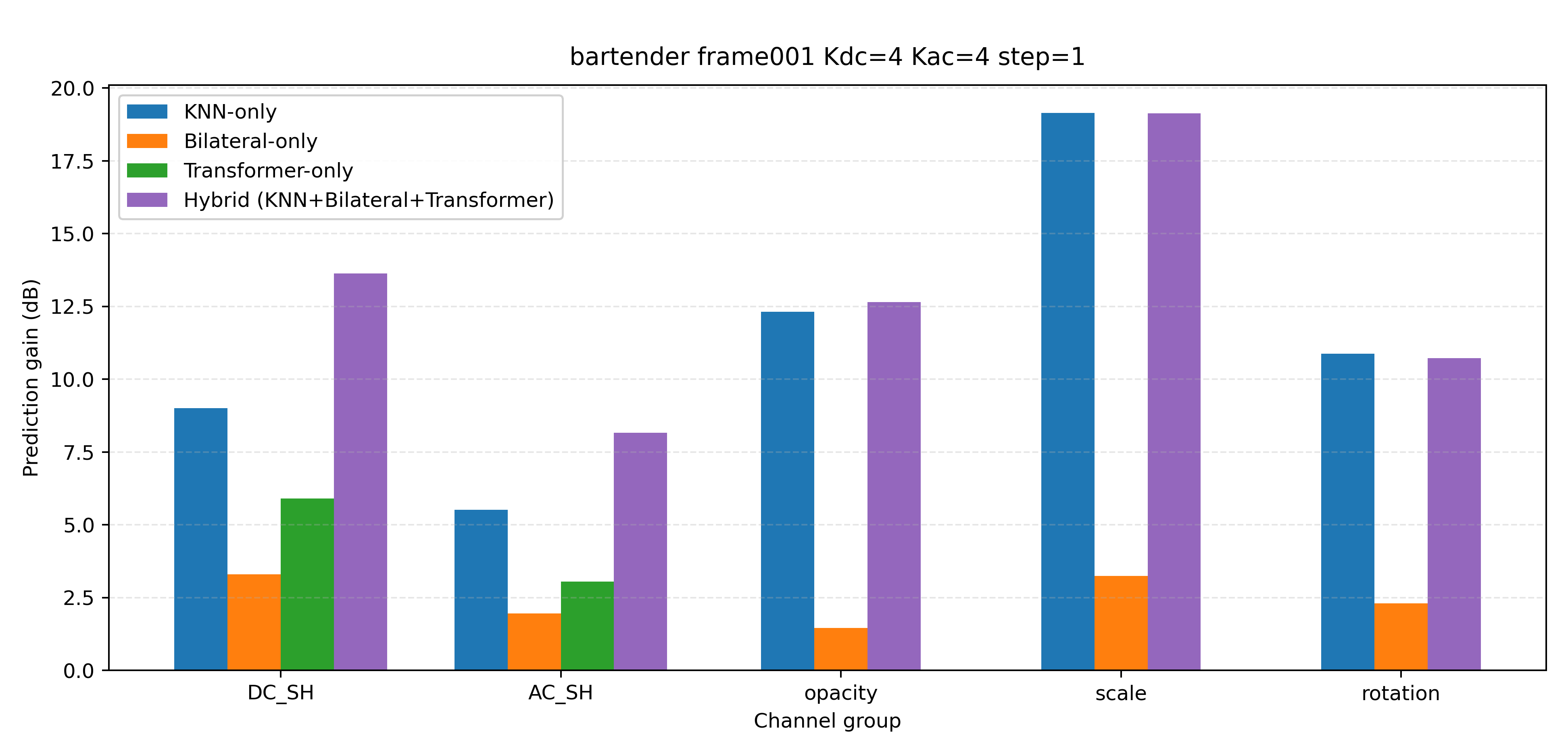}

        \vspace{-1mm}
        {\footnotesize (b) Predictor gain}
    \end{minipage}

    \vspace{-3mm}
    \caption{Rate--distortion and predictor analysis of InterGS+.
    (a) BD-rate reduction and BD-PSNR gain relative to the evaluated baselines.
    (b) Prediction gain of the individual predictor families and the adaptive
    hybrid scheme on a representative frame.}
    \label{fig:bd_predictor}
    \vspace{-6mm}
\end{figure}

% The BD results in Fig.~\ref{fig:bd_predictor}(a) confirm that the gains are not
% limited to the G-PCC anchor. InterGS+ also improves upon InterGS-Lite on
% all evaluated sequences in the current comparison, demonstrating that the
% learned SH predictor and representation-aware residual coding contribute
% additional compression efficiency beyond the existing inter-frame framework.
% The improvements on sequences unseen during training further indicate that the
% learned predictor generalizes beyond its training frames and sequence
% representations.

\begin{figure}[t]
    \centering

    \begin{minipage}[c]{0.28\linewidth}
        \centering
        \includegraphics[width=\linewidth]
        {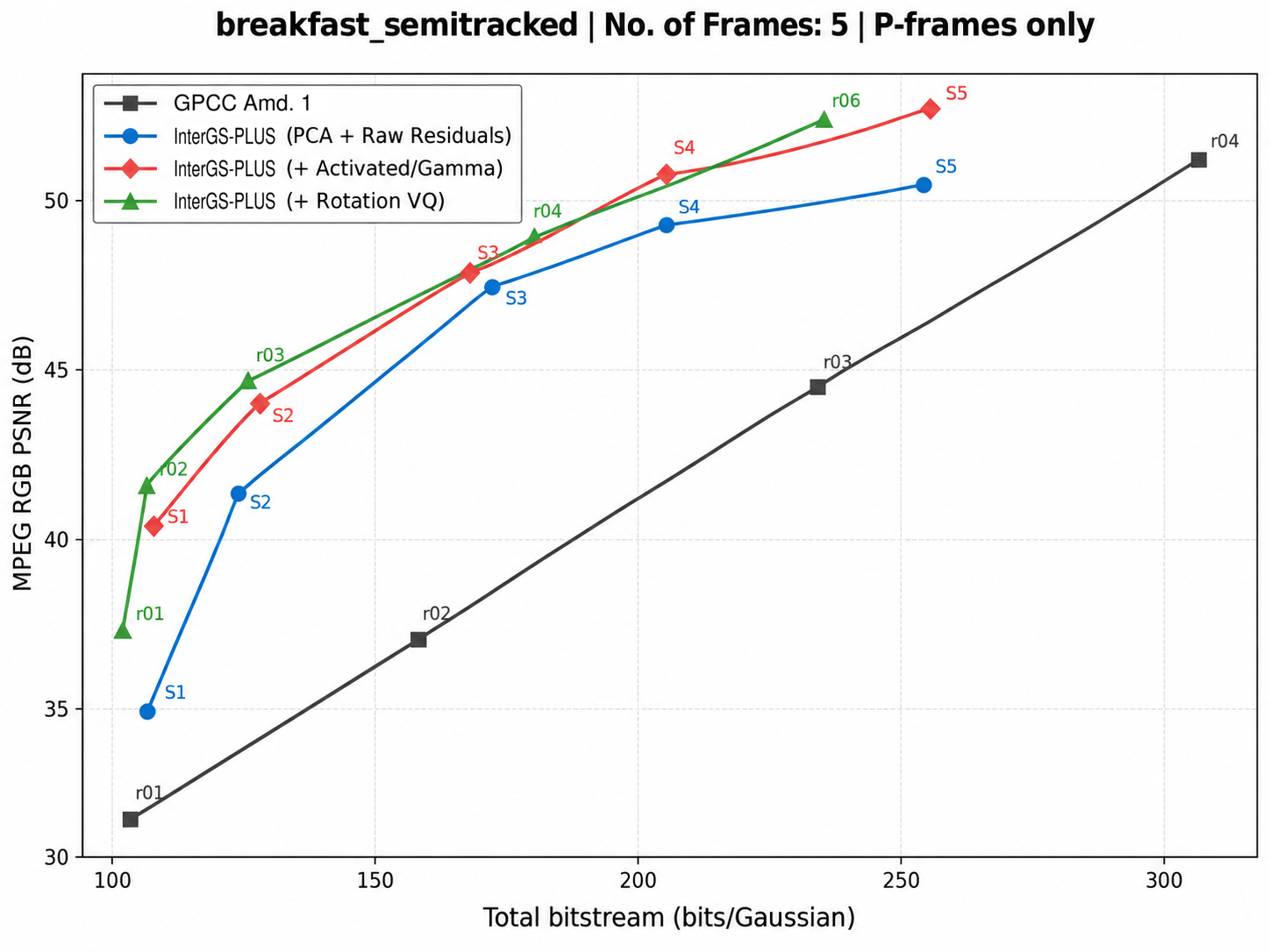}
    \end{minipage}
    \hfill
    \begin{minipage}[c]{0.28\linewidth}
        \centering
        \includegraphics[width=\linewidth]
        {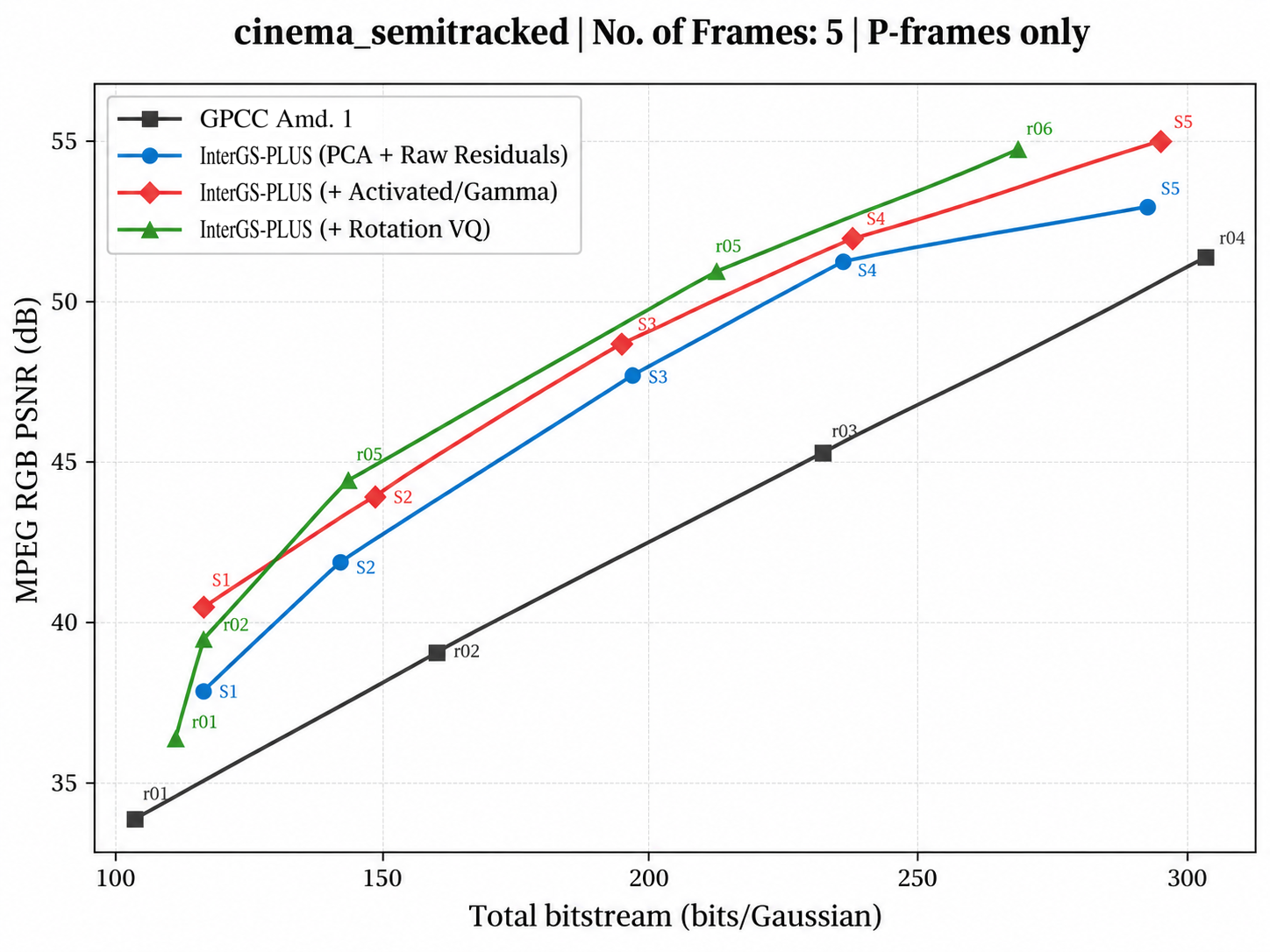}
    \end{minipage}
    \hfill
    \begin{minipage}[c]{0.28\linewidth}
        \centering
        \includegraphics[width=\linewidth]
        {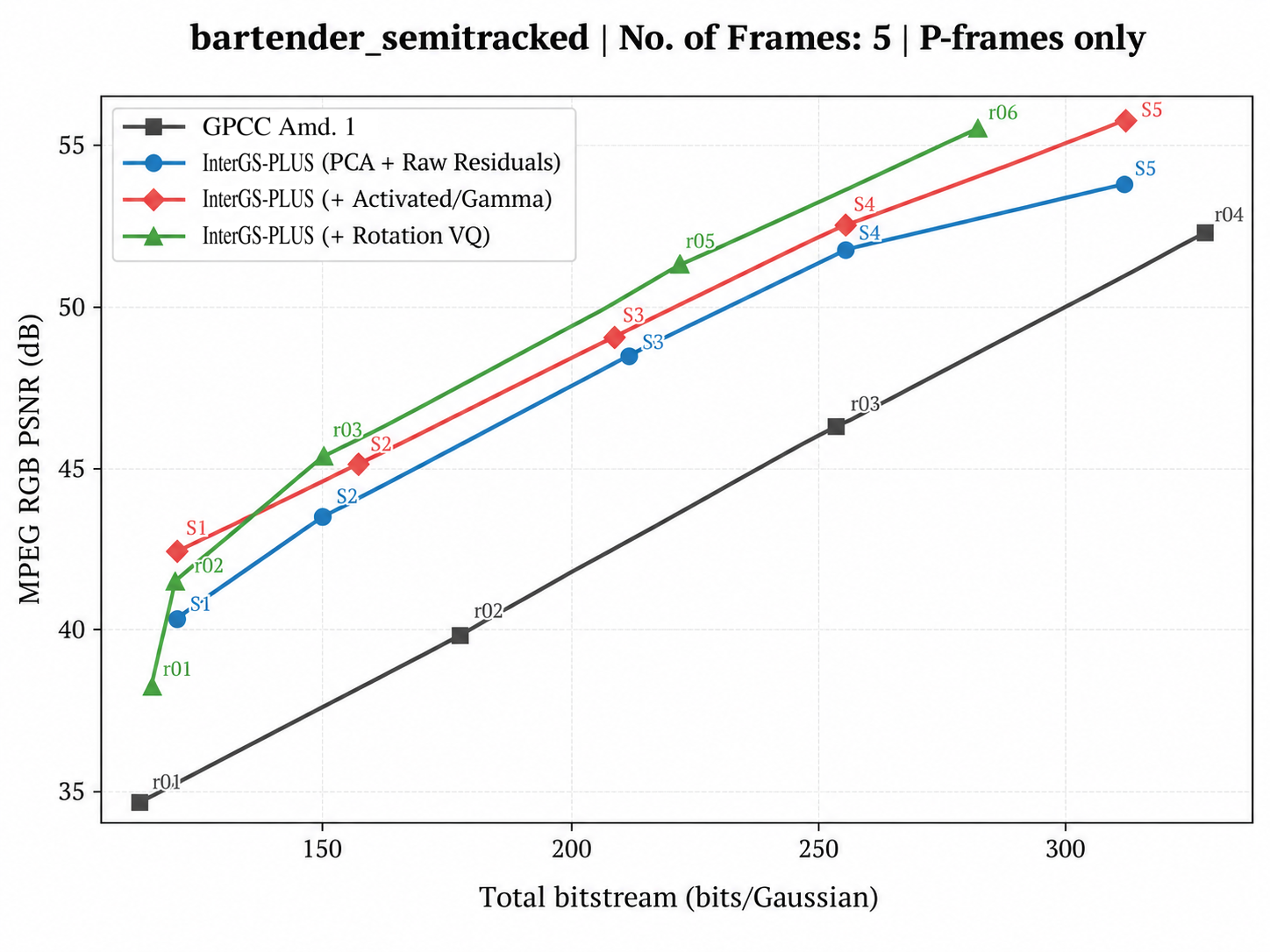}
    \end{minipage}

    \vspace{-1mm}
    \caption{Residual-coding ablation on the five P-frames of the semi-tracked
    \emph{Bartender}, \emph{Cinema}, and \emph{Breakfast} sequences.}
    \label{fig:residual_ablation}
    \vspace{-3mm}
\end{figure}

\vspace{-6mm}
\paragraph{Residual-coding ablation.}
% Figure~\ref{fig:residual_ablation} evaluates the five P-frames of the
% semi-tracked \emph{Bartender}, \emph{Cinema}, and \emph{Breakfast} sequences
% with the prediction configuration fixed across all InterGS+ variants.
% Activated gamma-companded residual coding consistently improves upon raw-domain
% coding, while quaternion rotation VQ provides additional gains over most of the
% operating range. This confirms that matching the residual representation to the
% underlying Gaussian attribute structure improves coding efficiency beyond
% temporal prediction alone.
% \noindent\textbf{Residual-coding ablation.}
 Figure~\ref{fig:residual_ablation} evaluates the five P-frames of the semi-tracked
Bartender, Cinema, and Breakfast sequences with the prediction configuration
fixed across all InterGS+ variants. Activated gamma-companded residual coding
consistently improves upon raw-domain coding, while quaternion rotation VQ
provides additional gains over most of the operating range, confirming that
attribute-matched residual representations improve coding efficiency beyond
temporal prediction alone.

\vspace{-5mm}
\paragraph{Predictor Analysis.}
% Figure~\ref{fig:bd_predictor}(b) shows that predictor effectiveness is
% attribute dependent. Lightweight KNN and bilateral predictors remain effective
% for non-SH attributes, whereas SH benefits more from the geometry-aware
% transformer and adaptive hybrid scheme. The transformer is nevertheless not
% optimal for every Gaussian: on the representative frame, it gives lower SH
% prediction error for $71.43\%$ of Gaussians, while bilateral prediction is
% preferred for the remaining $28.57\%$. This complementary behavior supports
% the per-Gaussian predictor selection used in InterGS+.
Figure~\ref{fig:bd_predictor}(b) shows that predictor effectiveness is
attribute dependent: KNN/bilateral prediction remains effective for non-SH
attributes, whereas SH benefits more from the transformer and adaptive hybrid
scheme. On the representative frame, the transformer yields lower SH prediction
error for 71.43\% of Gaussians, while bilateral prediction is preferred for
28.57\%, supporting per-Gaussian predictor selection.
% Table~\ref{tab:full_bpp_distribution} reports the contribution of each
% bitstream component at the highest-rate operating point; the final two rows
% give the total bitrate and corresponding PSNR for each sequence. InterGS+
% remains lightweight at decoding: learned inference is restricted to the SH
% stream, and the transformer contains only $53{,}964$ fixed parameters. Runtime
% was measured on an Intel Core i9-11900F CPU (8 cores/16 threads), 62\,GiB RAM,
% and an NVIDIA RTX 3090 GPU (24\,GB), running Ubuntu 22.04. Across the 10
% evaluated sequences, the average decoder runtime is
% \textbf{0.049 ms/Gaussian}.
Table~\ref{tab:full_bpp_distribution} reports the bitstream-component distribution at the
highest-rate operating point, with total bitrate and PSNR in the final rows.
InterGS+ remains lightweight at decoding: learned inference is confined to SH
and uses only 53,964 fixed transformer parameters. Complexity was measured on
an Intel Core i9-11900F CPU (8 cores/16 threads), 62 GiB RAM, and an NVIDIA
RTX 3090 GPU (24 GB) under Ubuntu 22.04. The reported timing covers decoding
only, excluding rendering, and averages 0.049 ms/Gaussian over the 10 sequences.
% Table~\ref{tab:full_bpp_distribution} reports the contribution of each
% bitstream component at the highest-rate operating point; the final two rows
% give the total bitrate and corresponding PSNR for each sequence. InterGS+
% remains lightweight at decoding: learned inference is restricted to the SH
% stream, and the transformer contains only $53{,}964$ fixed parameters. Complexity
% was measured on an Intel Core i9-11900F CPU (8 cores/16 threads), 62\,GiB RAM,
% and an NVIDIA RTX 3090 GPU (24\,GB), running Ubuntu 22.04. The reported timing
% covers decoding only and excludes rendering. Across the 10 evaluated sequences,
% the average decoding complexity is 0.049 ms/Gaussian.
% ============================================================
% Bitstream distribution
% ============================================================

\begin{table*}[t]
\vspace{-3mm}
\centering
\caption{Per-sequence bitstream component distribution at the highest-rate
operating point. All bitrate values are in bits per Gaussian (bpp).}
\label{tab:full_bpp_distribution}
\scriptsize
\resizebox{\textwidth}{!}{%
\begin{tabular}{lrrrrrrrrrr}
\toprule
\textbf{Bitstream Component} &
\textbf{Bartender ST} &
\textbf{Bartender T} &
\textbf{Breakfast ST} &
\textbf{Breakfast T} &
\textbf{Cinema ST} &
\textbf{Cinema T} &
\textbf{Makeup} &
\textbf{ManWithFruit T} &
\textbf{Musical} &
\textbf{Trio} \\
\midrule

DC SH
& 4.415 & 4.344 & 4.158 & 4.136 & 4.122 & 4.087 & 0.809 & 0.912 & 1.228 & 1.292 \\

AC SH
& 1.687 & 0.830 & 1.495 & 0.914 & 1.492 & 0.971 & 0.646 & 2.205 & 1.170 & 0.654 \\

Opacity
& 7.080 & 6.580 & 6.139 & 5.746 & 6.153 & 5.976 & 0.870 & 8.394 & 1.240 & 0.898 \\

Scale
& 19.949 & 18.470 & 18.905 & 16.695 & 21.225 & 20.125 & 3.915 & 9.847 & 4.654 & 3.764 \\

Rotation
& 4.010 & 3.415 & 3.593 & 3.070 & 3.811 & 3.121 & 0.818 & 2.102 & 0.763 & 0.761 \\

Decoder Metadata
& 0.478 & 0.661 & 0.707 & 0.710 & 0.550 & 0.886 & 0.468 & 8.312 & 0.624 & 0.468 \\

DC SH VQ Refinement
& 19.459 & 17.292 & 16.296 & 14.523 & 18.411 & 17.147 & 2.563 & 4.449 & 1.997 & 2.539 \\

AC SH VQ Refinement
& 195.386 & 146.775 & 140.842 & 105.883 & 178.474 & 137.962 & 47.535 & 140.723 & 38.994 & 51.074 \\

Codebook 1
& 0.011 & 0.015 & 0.016 & 0.016 & 0.012 & 0.020 & 0.010 & 0.190 & 0.014 & 0.010 \\

Codebook 2
& 0.024 & 0.033 & 0.035 & 0.036 & 0.028 & 0.044 & 0.023 & 0.409 & 0.031 & 0.023 \\

Rotation Codebook
& 0.011 & 0.015 & 0.016 & 0.016 & 0.012 & 0.020 & 0.011 & 0.195 & 0.014 & 0.011 \\

Rotation VQ Refinement
& 27.659 & 25.164 & 25.991 & 23.287 & 26.438 & 24.690 & 4.586 & 20.482 & 4.298 & 4.196 \\

Predictor Choice
& 5.124 & 4.844 & 4.120 & 3.597 & 4.957 & 4.667 & 1.630 & 3.086 & 1.521 & 1.431 \\

Geometry
& 34.272 & 22.872 & 29.685 & 29.665 & 34.453 & 23.333 & 26.749 & 39.838 & 28.255 & 27.293 \\

\midrule

\textbf{Total BPP}
& \textbf{319.567}
& \textbf{251.308}
& \textbf{251.997}
& \textbf{208.294}
& \textbf{300.136}
& \textbf{243.048}
& \textbf{90.636}
& \textbf{241.144}
& \textbf{84.805}
& \textbf{94.415} \\

\textbf{PSNR (dB)}
& \textbf{57.006}
& \textbf{53.425}
& \textbf{53.588}
& \textbf{54.207}
& \textbf{56.481}
& \textbf{52.533}
& \textbf{48.107}
& \textbf{64.858}
& \textbf{52.585}
& \textbf{47.323} \\

\bottomrule
\end{tabular}%
}
\vspace{-4mm}
\end{table*}

\vspace{-7mm}
\section{Conclusion}
\label{sec:conclusion}
\vspace{-4mm}
% We presented \emph{InterGS+}, a post-hoc inter-frame codec for dynamic 3D
% Gaussian Splatting that improves both temporal prediction and residual
% representation while preserving lightweight decoding. InterGS+ introduces a
% shared PCA domain for SH appearance, augments the InterGS-Lite predictor bank
% with a geometry-aware attention-based SH predictor, and forms non-SH residuals
% after accounting for renderer-domain nonlinearities and representation
% ambiguities. Canonicalized quaternion innovations are further coded jointly
% using vector quantization with scalar refinement. Experiments on 10 dynamic
% 3DGS sequences, including seven sequences unseen during predictor training,
% demonstrate consistent rate--distortion gains over GPCC and InterGS-Lite while
% retaining low decoding complexity. These results show that effective dynamic
% 3DGS compression benefits not only from stronger temporal prediction, but also
% from choosing prediction and residual domains that respect the structure of
% the underlying Gaussian attributes.

% Future work will investigate longer temporal dependencies and multi-reference
% prediction while retaining the lightweight post-hoc coding framework.
We presented \emph{InterGS+}, a lightweight post-hoc inter-frame codec for
dynamic 3DGS that combines a shared SH PCA domain, geometry-aware transformer
prediction, representation-aware non-SH residual coding, and VQ-based coding of
canonicalized quaternion innovations. Across 10 dynamic 3DGS sequences,
including seven unseen during training, InterGS+ consistently outperforms
G-PCC Amd.~1 and InterGS-Lite while maintaining low decoding complexity.
These results demonstrate the benefit of jointly improving temporal prediction
and attribute-aware residual representation. Future work will consider longer
temporal dependencies and multi-reference prediction within the same post-hoc
framework.
% We presented \emph{InterGS+}, a lightweight post-hoc inter-frame codec for
% dynamic 3DGS that improves both temporal prediction and residual coding.
% InterGS+ introduces a shared PCA domain for SH appearance, geometry-aware
% transformer-based temporal SH prediction, representation-aware non-SH residual
% formation, and joint VQ-based coding of canonicalized quaternion innovations. Experiments on 10 dynamic 3DGS sequences, including seven unseen during training, show consistent rate--distortion gains over GPCC and InterGS-Lite while maintaining low decoding complexity. The results highlight the benefit of combining stronger temporal prediction with attribute-aware residual representations. Future work will explore longer temporal dependencies and multi-reference prediction within the same lightweight post-hoc framework.

\vspace{-3mm}
\Section{References}
\vspace{-2mm}
\bibliographystyle{IEEEtran}
\bibliography{refs}

\end{document}